\documentclass{mecbic}
\providecommand{\anno}{2011} \providecommand{\firstpage}{1}
\providecommand{\lastpage}{2}
\usepackage{breakurl}

\usepackage{latexsym, amssymb}
\usepackage{breakurl}

\begin{document}

\title{Biologically Inspired Process Calculi, Petri Nets\\ and Membrane Computing}
\author{Gabriel Ciobanu
\institute{Romanian Academy, Institute of Computer Science \\
Ia\c si, Romania}
\email{gabriel@iit.tuiasi.ro}
}

\def\titlerunning{Process Calculi, Petri Nets and Membrane Computing}
\def\authorrunning{Gabriel Ciobanu}

\maketitle
\pagestyle{plain}
\pagenumbering{arabic}
\setcounter{page}{1}

\begin{abstract}

This volume represents the proceedings of the 5th Workshop on Membrane
Computing and Biologically Inspired Process Calculi (MeCBIC 2011), held
together with the 12th International Conference on Membrane Computing on
23rd August 2011 in Fontainebleau, France.

\end{abstract}

\medskip

MeCBIC is usually devoted to membrane computing and biologically inspired
process calculi (ambients, brane calculi). This year we also attracted papers
dealing with bio-inspired Petri nets in order to promote collaboration between
the Petri nets and membrane computing communities.

\smallskip

Biological membranes play a fundamental role in the complex reactions
which take place in cells of living organisms. Membrane systems were
introduced as a class of distributed parallel computing devices inspired
by the observation that any biological system is a complex hierarchical
structure, with a flow of biochemical substances and information that
underlies their functioning. The modelling and analysis of biological
systems has also attracted considerable interest from both the Petri nets and
the process calculi research communities.
A deeper investigation of the relationships between these formalisms is
interesting, providing valuable cross fertilization of these research areas.
Membrane computing deals with the computational properties, making use
of automata, formal languages, and complexity results. Certain process
calculi, such as mobile ambients and brane calculi, work also with
notions of compartments and membranes. Petri nets are used to model and
analyze the biological systems.

\medskip

The submitted papers describe biologically inspired models and calculi,
biologically inspired languages, properties of biologically inspired
models and languages, theoretical links and comparison between different
models. All submitted papers were reviewed by three or four referees. We
thank the authors and reviewers for doing an excellent job; without
their enthusiastic work this volume would not have been possible.
We are indebted to the members of the Programme Committee:\\

\begin{tabular}{l@{\hspace{1ex}}l@{\hspace{6ex}}l@{\hspace{6ex}}l}
Bogdan Aman & Jean-Louis Giavitto & Gethin Norman & Jason Steggles\\
Roberto Barbut & S.N. Krishna & Andrew Phillips & Angelo Troina\\
Marco Bernardo & Jean Krivine & G. Michele Pinna & Sergey Verlan\\
Paola Bonizzoni & Paolo Milazzo & Franck Pommereau &  Gianluigi Zavattaro.\\
Gabriel Ciobanu (chair)\\
\end{tabular}
\smallskip

\noindent
We express our gratitude to the invited speakers Jetty Kleijn and Cosimo Laneve
for their very interesting talks. The first talk presents similarities and differences
between Petri nets and membrane systems, and how to enhance the Petri nets
in order to faithfully model the dynamics of the biological phenomena
represented by membrane systems and reaction systems.
The second talk deals with reversibility in massive concurrent systems;
reversible structures for massive concurrent systems are introduced and studied,
and an equivalence on computations that abstracts away from the order of
causally independent reductions is defined.

\medskip

The main aim of the workshop is to bring together researchers working on
membrane computing, in biologically inspired process calculi, and in
Petri nets in order to present their recent results and to discuss new
ideas concerning these formalisms, their properties and relationships.
Many thanks to Sergey Verlan and Maciej Koutny for their help in organizing
the workshop, and to Bogdan Aman for his help in preparing this volume.





\end{document}